# Matter with apparent and hidden spin physics

Jia-Xin Xiong[1,†], Xiuwen Zhang[1], Lin-Ding Yuan[2], and Alex Zunger[1,*]

[1]Renewable and Sustainable Energy Institute, University of Colorado, Boulder, 80309, USA

[2]Department of Materials Science and Engineering, Northwestern University, Evanston, Illinois 60208, USA

## SIGNIFICANCE

Discoveries in quantum physics of matter are often guided and judged according to previously expected enabling conditions of the said effect. For example, Rashba effect in spin physics is known to be enabled by broken inversion symmetry in nonmagnetic systems in the presence of spin-orbit coupling (SOC). A newly discovered material said to carry the 'apparent Rashba effect' is judged by following these three enabling conditions. In contrast, *hidden effects* refer to the general condition where the nominal global symmetry would disallow the effects, whereas the symmetry of local parts of the systems would allow them. For example, *hidden* Rashba effect can show spin polarization even if the global symmetry violates the required broken inversion symmetry, but the structure consists of sectors that are individually non-centrosymmetric. Such '*hidden effects'* were often dismissed as being '*mistaken effects'* due to some extrinsic sample imperfection, but they could indeed be intrinsic property pertaining to the perfect crystal. Examples of *spin-unrelated* hidden effects include hidden chirality, valley polarization, intrinsic circular polarization, piezoelectric or orbital polarization. This Perspective describes a unified framework of spin-related classification of apparent and hidden spin effects into four categories, taking a 'real materials' view with specific examples of compounds with experimental relevance. It further discusses the tunability and switching of apparent spin splitting and hidden spin polarization in antiferromagnets, pointing out future opportunities of research.

## SUMMARY

Materials with interesting physical properties are often designed based on our understanding of the target physical effects. The physical properties can be either explicitly observed ("apparent") or concealed by the perceived symmetry ("hidden") but still exist. Both are enabled by specific symmetries and induced by certain physical interactions. Using the underlying approach of condensed matter theory of *real materials* (rather than schematic model Hamiltonians), we discuss apparent and hidden physics in real materials focusing on the properties of spin splitting and spin polarization. Depending on the enabling symmetries and underlying physical interactions, we classify spin effects into four categories with each having two subtypes; representative materials are pointed out. We then discuss the electric tunability and switch of apparent and hidden spin splitting and polarization in antiferromagnets. Finally, we extend "hidden effects" to views that are *farsighted* in the sense of resolving the correct atomistic and reciprocal symmetry and replaced by the incorrect higher symmetry. This framework could guide and enable systematic discovery of such intriguing effects.

[†] Corresponding author: Jia-Xin.Xiong@colorado.edu
[*] Corresponding author: alex.zunger@colorado.edu

# I. Introduction

The understanding of enabling conditions of physical effect *X*—conditions without which effect *X* is not possible—are often used to rule in or out the identification of *X*. An instance of "*apparent effect*" *X* in the realm of *polarized optics* could include the optical circular polarized luminescence enabled by broken inversion symmetry (e.g., in transition-metal dichalcogenides[1]), or piezoelectric polarization and second harmonic generation in semiconductors[2–4]. Such enabling conditions can consist of (i) a basic description of allowed configurations, (ii) a symmetry principle, and (iii) a needed physical interaction term. For example, in the realm of *spin physics* the effect of *X*=Splitting of spin bands, as in the Rashba effect[5,6] or the Dresselhaus effect[7] is enabled, by (i) nonmagnetic configurations, having (ii) broken inversion symmetry and experiencing (iii) spin-orbit-coupling (SOC) interaction. The distinguishing feature (auxiliary symmetries) is that the Rashba effect needs polar symmetry whereas the Dresselhaus effect needs nonpolar symmetry.

Such apparent effects can sometimes transform into *hidden* (but nevertheless existing) effects when some of the enabling conditions are unavailable. Indeed, the removal of an enabling condition from the corresponding apparent effect, sets up what will be referred to as the "hidden effect". An example is where the nominal global symmetry would disallow effect *X*, whereas the local symmetry would allow *X*. This could occur, for example when a unit cell contains separate sectors, each having their own local broken inversion symmetry, but the sectors taken together can be mutually compensating, leading to effective centrosymmetric global symmetry. Those different symmetries naturally allow different physical effects. For example, the observation of anisotropic optical circular polarized emission in centrosymmetric (even-layered) transition-metal dichalcogenide compounds[8] or the observation of piezoelectric polarization in nominally centrosymmetric matter[9] can be considered as hidden effects. Such '*hidden effects'* were often dismissed as being '*mistaken effects'* due to some extrinsic sample imperfection, but they could indeed be intrinsic property pertaining to the perfect crystal. Another example of hidden effect is the recent report of differential absorption of circularly polarized light in a centrosymmetric crystalline solid, which was thought to only exist in structures in absence of mirror or inversion symmetry[10]. Huang et al. revealed the hidden chirality in $NaCu_5S_3$[11]: The parent phase is non-chiral with the space group of $P6_3/mcm$, thus not suggesting this compound relevant to chiral physics, yet its non-chiral phase is made of two chiral sublattices with the space group of $P6_322$; The competition between these two chiral sublattices determines either the non-chiral phase ($P6_3/mcm$; hidden chirality) or the ground chiral phase ($P6_322$). Other spin-unrelated effects that can be considered as hidden include hidden valley polarization and intrinsic circular polarization[8], hidden piezoelectric polarization[12], and hidden orbital polarization[13] in different centrosymmetric systems. In these spin-unrelated hidden effects, the local inversion asymmetry enables valley polarization and local electric dipoles.

In the current article we focus on apparent vs hidden effects within the realm of quantum matter where the electron spin plays a central role. Examples of hidden spin effects arising from removal of enabling conditions are illustrated in Fig. 1. We focus on two physical properties here, i.e., the splitting of the energy bands ("spin splitting") and the polarized states with opposite spin components ("spin polarization") in the reciprocal space. When one removes the required enabling broken inversion symmetry [indicated in Fig. 1(B) by crossing out the operation of broken $I$ or $\Theta I$, where $I$ is spatial inversion, $\Theta$ is time reversal] from *X*=apparent SOC-induced spin-splitting (Category A). The apparent effect cannot exist globally, yet a hidden version of the effect, *X*=hidden SOC-induced spin polarization (Category B), can exist locally. An example is the nonmagnetic $BaNiS_2$[14–16], which has hidden Rashba spin polarization induced by non-centrosymmetric local sectors. Similarly, regarding the effect *X*=apparent SOC-independent spin splitting systems (Category C) where the required enabling condition of SOC is removed [indicated in Fig. 1(C) by crossing out the "SOC"], and the condition of time- and spatial-reversal symmetry is no longer broken [indicated in Fig. 1(D) by crossing out the operation of broken $\Theta I$ and $UT$, here $U$ is a spin rotation that reverses the spin ordering, $T$ is a fractional translation], the apparent effect (C) vanishes, but its hidden counterparts (Category D) of SOC-independent spin polarization can remain. An example is the antiferromagnetic (AFM) $Ca_2MnO_4$[17], which shows Zeeman-type local spin polarization induced by local magnetic sectors. In these cases, the apparent spin *splitting* is converted to hidden spin *polarization* since in the latter case the global symmetry protects spin degeneracy, i.e., no spin splitting.

In addition, there are 'auxiliary symmetries' that do not influence the very existence of spin splitting or spin polarization but shape the properties of spin polarization. For example, the auxiliary symmetry of presence of polar sectors in Rashba systems creates a local helical spin polarization or spin texture [Category (B,1), as in $LaOBiS_2$[12]]. Also, the auxiliary symmetry of presence of non-centrosymmetric non-polar sectors in Dresselhaus systems leads to local spin polarization [Category (B,1), as in NaCaBi[12]]. For the AFM compounds, the auxiliary symmetry of spin interconversion and polarity leads to various spin textures of subgroups [item (C,1), as in $MnF_2$[18], MnTe[19,20], $Mn_4Nb_2O_9$[21], and $BiCrO_3$[21]].

***(A): Apparent SOC-induced spin splitting***

| (A,1) | (A,2) |
|---|---|
| • Broken $I$<br>• SOC<br>• NM | • Broken $\Theta I$<br>• SOC<br>• AFM background |

***(B): Hidden SOC-induced spin polarization***

| (B,1) | (B,2) |
|---|---|
| • ~~Broken $I$~~<br>• SOC<br>• NM | • ~~Broken $\Theta I$~~<br>• SOC<br>• AFM background |

***(C): Apparent SOC-independent spin splitting***

| (C,1) | (C,2) |
|---|---|
| • Broken $\Theta I$ and $UT$<br>• ~~SOC~~<br>• AFM | • Broken $\Theta I$ and $UT$<br>• ~~SOC~~<br>• FM |

***(D): Hidden SOC-independent spin polarization***

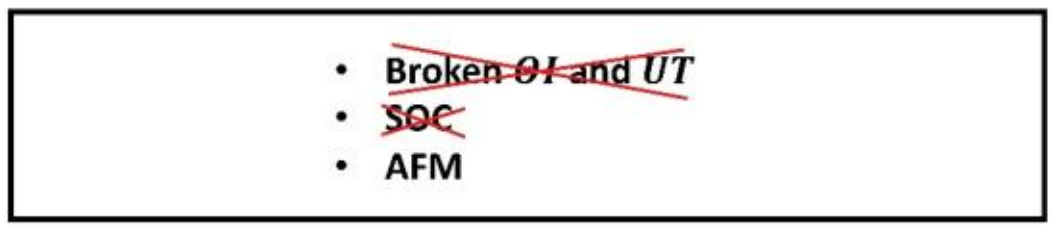

**Figure 1:** Enabling symmetries of spin-related apparent and hidden effects that are dependent or independent of SOC. "Broken $I$" means that the inversion symmetry is broken, whereas "broken $I$" with a red cross means preserving inversion symmetry. The notations are time-reversal symmetry $\Theta$, inversion symmetry $I$, spin-rotation *SU(2)* symmetry $U$, and fractional translation symmetry $T$. Abbreviations are: "SOC" for spin-orbit coupling, "NM" for nonmagnet, "AFM" for antiferromagnet, and "FM" for ferromagnet.

Theoretical predictions and experiments on hidden spin polarization in non-magnetic centrosymmetric crystals have spurred research into other physical effects that are normally forbidden under apparent global crystal symmetry. These include hidden field-induced non-reciprocal transport induced by hidden spin polarization[22] and hidden spin polarization in Bi-based cuprate superconductors[23] in centrosymmetric systems. In these hidden spin effects, the local inversion asymmetry enables the local spin polarization that can be coupled with the antiferromagnetic order or coexist with superconductivity.

Another situation that leads to hidden effects is when a low-resolution approximate theory or simplified experiment captures a false high average symmetry instead of the actual lower symmetry. The latter case can predict different symmetry-controlled physical effects hidden by the higher false symmetry. An example of low-resolution theory is when a small basis set is used to expand a as complex wavefunction as in small basis $k \cdot p$ in nanoscience. Sec. VIII provides some examples of effects hidden by such 'farsightedness'.

Understanding what is behind the apparent and hidden effect is crucial as this explains seemingly paradoxical phenomena that appear hidden by global symmetry. This also significantly broadens the material pools for target properties to be designed in. The present perspective article will emphasize (a) the main role of enabling symmetry, (b) the required physical interaction, (c) how enabling effects can be decorated by auxiliary conditions that shape properties of the effect (but not changing its existence), and (d) how the structure of this theory is connected to experimental observations and clarification of what is apparent and what is hidden.

Our goal is to rationalize the currently known hidden effects vis their apparent effect counterpart, with the focus on apparent spin splitting and hidden spin polarization. This Perspective attempts to establish a general understanding of hidden effects that are disallowed by global symmetry but can nevertheless exist because of local symmetries. Each apparent effect can have multiple corresponding hidden effects depending on the global symmetry that compensates the local physical properties $X_p$. This richness of apparent effects lies at the basis of promoting deeper understanding of the physics of hidden effects.

In the next sections, we classify apparent and hidden spin physics first in nonmagnets (mainly discussing Rashba and Dresselhaus effects and their hidden counterparts) and then expand the scope to magnets.

# II. Classification of apparent spin splitting and hidden spin polarization

We start by building the classification of Fig. 2—the driving schematic of this paper—in three steps. First ,we define the "spin-splitting types" (SST) of Table I; Second, we use the different SSTs' to discuss two switches of enabling symmetry and required physical interactions and obtain the categories (A, B, C, D); Third, we point out that each category splits into two subtypes (denoted as subtype 1 and 2), depending on different magnetic configurations in either global or local environments of the systems.

*Using magnetic space group or spin space group to classify spin-split and spin-degenerate magnetic systems:* We have used the crystallographic space group to describe the relativistic-spin-splitting-enabling symmetries for nonmagnetic systems. For the description of spin-splitting effect in collinear magnetic systems, in the limit of no SOC, we used $\Theta I$ and $UT$ symmetries. This approach is general and aligns with the spin space group analysis[24–26]. Following the notation of Daniel Litvin et al.[27,28] The absence of the two symmetries implies no inversion or spatial translation connecting the atomic sites with opposite magnetic moments while keeping the crystal structure intact. Because there is $\Theta U$ symmetry in collinear magnets, the existence of the $UT$ symmetry in the spin space group implies there is a $\Theta T$ symmetry in the magnetic space group (by definition, this corresponds to magnetic space group type IV). Therefore, the magnetic space group is equivalent to the spin space group in terms of describing non-relativistic spin splitting (NRSS) effect in collinear magnets (Ref.[17] second paragraph of Discussion section discusses explicitly the equivalence of magnetic space group and spin space group in terms of describing NRSS). As such, in this Perspective, we will use the magnetic space group to describe the enabling symmetry of collinear magnetic systems.

In the first step of constructing the classification of Fig.2 , we assign "spin-splitting types" (SST)[29], as summarized in Table I, based on their spin-splitting enabling symmetries, required physical interactions, and when appropriate, magnetic configurations:

**Table I:** Spin-splitting types (SST) based on enabling symmetries, required physical interactions for apparent spin splitting or hidden spin polarization, and magnetic configurations. The notations are time-reversal symmetry $\Theta$, inversion symmetry $I$, spin-rotation (belonging to the *SU(2) spin ½ group)* that reverses the collinear/coplanar spin order $U$, and fractional translation symmetry $T$. Abbreviations are: "SOC" for spin-orbit coupling; "NM" for nonmagnet, "AFM" for antiferromagnet, "FM" for ferromagnet. The definition is from Ref.[21,29]. The $UT$ symmetry is present in nonmagnets without SOC; however, since SOC is included in SST-0 and SST-6, $UT$ symmetry need not be included for them. The second column provides a short, simplified reference name for each SST.

| **Spin-splitting type** | **Intuitive name** | **Enabling symmetry** | | **Physical interaction** |
|---|---|---|---|---|
| | | **Having $\Theta I$** | **Having $UT$** | |
| SST-0 | Rashba/Dresselhaus spin splitting in NM | No $I$ | -- | SOC required |
| SST-1 | Hidden spin polarization in AFM without $UT$ symmetry | Yes | No | Magnetism required. SOC not required |
| SST-2 | Hidden spin polarization in AFM having $UT$ symmetry | Yes | Yes | SOC or magnetism required |
| SST-3 | Rashba/Dresselhaus spin splitting in AFM having $UT$ symmetry | No | Yes | SOC required. AFM as background |
| SST-4 | Non-relativistic spin splitting in AFM without $UT$ symmetry | No | No | Magnetism required. SOC not required |
| SST-5 | Zeeman spin splitting in FM without $UT$ symmetry | No | No | Magnetism required. SOC not required |
| SST-6 | Hidden Rashba/Dresselhaus spin polarization in NM | Having $I$ | -- | SOC required |

SST-0 [1]: Rashba/Dresselhaus spin splitting in nonmagnets without inversion symmetry, exemplified by rhombohedral GeTe.

SST-1: Hidden spin polarization in antiferromagnets without $UT$ symmetry, exemplified by tetragonal CuMnAs.

[1] Also referred to as "SST-7" in Ref.[29] but here we use "SST-0" as the first prototype as this type (including Rashba effect) is well known to the community.

SST-2: Hidden spin polarization in antiferromagnets having $UT$ symmetry, exemplified by rocksalt NiO.

SST-3: Rashba/Dresselhaus spin splitting in antiferromagnets having $UT$ symmetry, exemplified by tetragonal $BiCoO_3$.

SST-4: Non-relativistic spin splitting in antiferromagnets without $UT$, exemplified by rutile $MnF_2$.

SST-5: Zeeman spin splitting in ferromagnets without $UT$ symmetry, exemplified by cubic $Ni_2FeGa$.

SST-6: Hidden Rashba/Dresselhaus spin polarization in nonmagnets having inversion symmetry, exemplified by tetragonal $BaNiS_2$.

The recently termed "altermagnet"[30,31] is a subgroup of the previously established SST-4 AFM category with ($\Theta I$, $UT$)=(No, No) where the two spin sublattices are connected by spin-interconverting *rotational* symmetry. A more detailed discussion of the relationship between "altermagnets" and SST-4 AFM compounds is given in Sec. V [Category (C, 1)].

*In the second step*, we group the seven SSTs into four categories A, B, C, D shown in the first column of Fig. 2, classifying the spin-related physical effects X that are enabling symmetry $X_s$ and induced by the physical interactions $X_i$ according to the questions (i) and (ii) below:

(i) Is the *enabling symmetry* $X_s$ broken in the global system? This is denoted as $X_s$=Yes, corresponding to SST-0, 3, 4, 5. Or, is $X_s$ broken only in the local symmetry but not broken in the global symmetry? This is denoted as $X_s$=No, corresponding to SST-1, 2, 6.

(ii) Is the *physical interaction* $X_i$ of SOC required (SOC-induced) or not (SOC-independent)? They are denoted as $X_i$=Yes, corresponding to SST-0, 3, 6, and $X_i$=No, corresponding SST-1, 2, 4, 5, respectively.

| Type of effect $X$ | Name of effect and example | Conditions | | | | Results |
|---|---|---|---|---|---|---|
| | | Spin-splitting prototype | Enabling symmetry ($X_s$) | | Are global and local enabling symmetries $X_s$ broken? | Does spin splitting exist at generic $k$-points when SOC is neglected or included? |
| | | | Is $\Theta I$ preserved? | Is $UT$ preserved? | | |
| (A) Apparent SOC-induced spin splitting | 1. Spin splitting (R1/D1) in NM : non-CS with SOC (**BiTeI[a]; GaAs[b]**) | SST-0 | No ($S_I = I$) | -- | (Yes, Yes) | (No; Yes) |
| | 2. Spin splitting in background AFM: non-CS; with SOC (**$BiCoO_3$[c]**) | SST-3 | No | Yes | (Yes, Yes) | (No; Yes) |
| (B) Hidden SOC-induced spin polarization | 1. Hidden spin polarization (R2/D2) in NM : CS; with SOC (**$LaOBiS_2$[d,e]; NaCaBi[e]**) | SST-6 | Yes ($S_I = I$) | -- | (No, Yes) | (No; No) |
| | 2. Hidden spin polarization in background AFM: CS; with SOC | SST-2 | Yes | Yes | (No, Yes) | (No; No) |
| (C) Apparent SOC-independent spin splitting | 1. Non-relativistic spin splitting in AFM: no SOC (**$MnF_2$[f], MnTe[g]**) | SST-4 | No | No | (Yes, Yes) | (Yes; Yes) |
| | 2. Zeeman spin splitting in FM: no SOC (**$Ni_2FeGa$[h]**) | SST-5 | No | No | (Yes, Yes) | (Yes; Yes) |
| (D) Hidden SOC-independent spin polarization | 1. Hidden non-relativistic spin polarization in AFM: no SOC; AFM local sectors (**$Ca_2MnO_4$[i], $MnS_2$[j]**) | SST-1 | Yes | No | (No, Yes) | (No; No) |
| | | SST-2 | Yes | Yes | (No, Yes) | (No; No) |
| | | SST-3 | No | Yes | (No, Yes) | (No; Yes) |
| | 2. Hidden non-relativistic spin polarization in AFM: no SOC; FM local sectors (**CuMnAs[j], $FeCl_2$[j]**) | SST-1 | Yes | No | (No, Yes) | (No; No) |
| | | SST-2 | Yes | Yes | (No, Yes) | (No; No) |
| | | SST-3 | No | Yes | (No, Yes) | (No; Yes) |

**Figure 2:** Four categories of apparent and hidden spin splitting and polarization with their enabling symmetry conditions and physical interactions. The notations are time-reversal symmetry $\Theta$, inversion symmetry $I$, spin-rotation that reverses the collinear spin order $U$, and fractional translation symmetry $T$. Note that for (A)-1 and (C)-1, $S_I = I$ is broken as $\Theta$ is present in non-magnets. Abbreviations in the second column are: "CS" for centrosymmetric, "SOC" for spin-orbit coupling, "NM" for non-magnet, "AFM" for antiferromagnet, "FM" for ferromagnet. The representative compounds (column 2 of the figure) are found in the following references: BiTeI: theoretical [a]Ref.[32]; GaAs: theoretical [b]Ref.[33]; $BiCoO_3$: theoretical [c]Ref.[34]; $LaOBiS_2$: theoretical [d]Ref.[12] and experimental [e]Ref.[35]; NaCaBi: theoretical [d]Ref.[12]; $MnF_2$: theoretical [f]Ref.[18]; MnTe: experimental [g]Ref.[20]; $Ni_2FeGa$: theoretical [h]Ref.[36]; $Ca_2MnO_4$: experimental [i]Ref.[37]; $MnS_2$: theoretical [j]Ref.[29]; CuMnAs, theoretical [j]Ref.[29]; $FeCl_2$: theoretical [j]Ref.[29].

*In the third step*, we combine either of the two options from (i) $X_s$=Yes or $X_s$=No with either of the two options of (ii) $X_i$=Yes (SOC-induced) or $X_i$=No (SOC-independent), obtaining four categories (A) Apparent SOC-induced spin splitting, (B) Hidden SOC-induced spin polarization, (C) Apparent SOC-independent spin splitting, and (D) Hidden SOC-independent spin polarization. Each category can be divided into subtype 1 or subtype 2, based on the magnetic systems of the effects in either global or local environments. This gives the complete classification of Fig. 2:

(A,1) Global and local SST-0, requiring SOC in NM compounds.

(A,2) Global and local SST-3, requiring SOC in AFM compounds.

(B,1) Global SST-6 and local SST-0, requiring SOC in NM compounds.

(B,2) Global SST-2 and local SST-3, requiring SOC in AFM compounds.

(C,1) Global and local SST-4, requiring magnetic order in AFM compounds.

(C,2) Global and local SST-5, requiring magnetic order in FM compounds.

(D,1) Global SST-1, or global SST-2, or global SST-3, and local SST-4 AFM sectors, requiring magnetic order in AFM compounds.

(D,2) Global SST-1, or SST-2, or SST-3, and local SST-5 FM sectors, and requiring magnetic order in AFM compounds.

In the rest of the paper, we will discuss the physics of each of these phenomena (A,1) to (D,2) including real material examples.

# III. Category (A) spin effects: Apparent SOC-induced spin splitting

Apparent SOC-induced spin splitting is the most familiar type of spin splitting physics, which is shown as Category (A) in Fig. 2. It is enabled by breaking the inversion symmetry $I$ (combined with time-reversal symmetry $\Theta$ in magnetic systems) and is induced by SOC, which can exist in both NM and AFM materials. Except for the enabling conditions, there are other auxiliary conditions, such as polarity and chirality, that shape the properties of spin polarization or spin texture at different wavevectors. This will be discussed in subsection (A,1).

## (A,1) Apparent SOC-induced spin splitting in non-centrosymmetric nonmagnetic materials

*The role of spin-orbit coupling in spin splitting:* An electron moving with momentum $p$ and mass $m$ in a crystal with potential gradient $\nabla V(r)$ enabled by inversion asymmetry gives rise to an effective magnetic field $B_{eff} \propto [\nabla V(r) \times p]$. The inversion asymmetry (violating $I$), as the enabling symmetry, leads to the coupling between electron spin $\sigma$ and orbital angular momentum $r \times p$, lifting the spin degeneracy in NM crystals that is originally protected by time-reversal symmetry $\Theta$ and inversion symmetry $I$ with each subband states being spin polarized. The splitting of energy bands is usually associated with spin polarization in each band, leading to the spin splitting with a characteristic spin texture in the reciprocal space. In addition to spin splitting, there is another splitting of energy bands, termed "band splitting" with vanishing spin polarizations in each split band, which is also driven by SOC in non-centrosymmetric nonmagnetic systems but appears in specific $k$-paths[38]. For example, the band splitting with vanishing spin polarization was predicted to appear in zinc-blende GaAs[38] along the $\Gamma$-$L$ path (i.e., the [111] direction), which is protected by the little group $C_{3v}$.

*Auxiliary symmetry that shapes the spin texture:* The SOC-induced spin splitting in non-centrosymmetric non-magnetic materials can be shaped by controlling additional, auxiliary symmetry conditions that do not enable the splitting, but control its properties and the spin texture into four principal groups: Rashba texture, Dresselhaus texture, Weyl texture and mixing of the above. When the point group symmetry around certain wavevectors is non-chiral[39] (i.e., the system is identical to its mirror image): The SOC-induced spin splitting and polarization can be divided into two types, i.e., Rashba[40,41] and Dresselhaus effect[7], according to different origins to break the inversion symmetry. The Rashba effect characterized by helical or tangential spin textures results from non-zero dipoles in polar point groups of $C_{3v}$, $C_{4v}$, and $C_{6v}$, whereas the Dresselhaus effect characterized by tangential-radial spin textures originates from non-polar point groups of $S_4$, $C_{3h}$, $D_{2d}$, $D_{3h}$, and $T_d$. Particularly, in polar point groups $C_s$ and $C_{2v}$, Rashba-type spin texture can coexist with Dresselhaus-type spin texture. Due to the external tunability, the Rashba effect is more appealing than the Dresselhaus effect for materials engineering.

The spin texture becomes more complex and changes to other types when the point group symmetry around certain wavevectors is chiral[39] (i.e., the system cannot be superposed on its mirror image): When the point group symmetry is non-polar as in $D_3$, $D_4$, $D_6$, $T$, and $O$, the spin texture will be radial as is the case in Weyl-type texture; when the point group symmetry is polar as in $C_3$, $C_4$, and $C_6$, the spin texture will have other patterns which allow the mixing of Rashba and Weyl-type spin textures. In the non-polar point group $D_2$, Dresselhaus-type spin texture can coexist with Weyl-type spin texture.

## Category (A): Apparent SOC-induced spin splitting

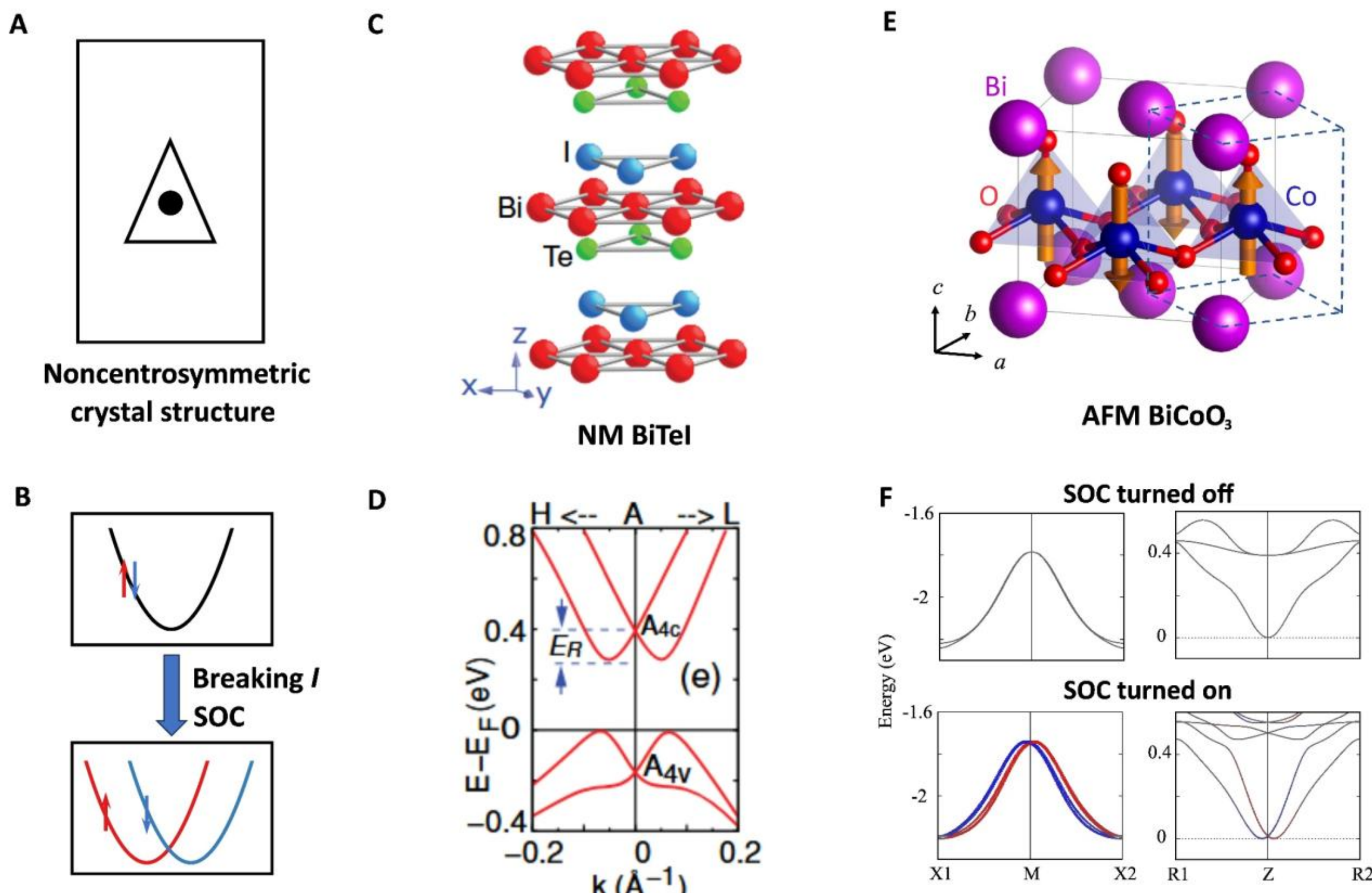

**Figure 3. Apparent SOC-induced spin splitting in nonmagnets [Category (A)].**

(A) Schematic of the crystal structure with inversion symmetry violated.

(B) Schematic of the band structure with SOC-induced spin splitting.

(C) Crystal structure of NM non-centrosymmetric trigonal BiTeI. Pink, blue, and green atoms denote Bi, Te, and I, respectively. Reproduced with permission of theoretical Ref.[32], copyright 2011, The American Physical Society.

(D) Band structure of NM non-centrosymmetric BiTeI with SOC-induced spin splitting. Reproduced with permission of theoretical Ref.[32], copyright 2011, The American Physical Society.

(E) Crystal structure of AFM non-centrosymmetric tetragonal $BiCoO_3$. Purple, blue, and red atoms denote Bi, Co, and O, respectively. Orange arrows represent the local magnetic moments. The dashed lines show the NM unit cell. Reproduced with permission of theoretical Ref.[34], copyright 2019, The American Physical Society.

(F) Band structure of VBM and CBM of AFM non-centrosymmetric $BiCoO_3$ with SOC-induced spin splitting. X1 (R1) and X2 (R2) points are equivalent high-symmetric k-points X (R) in different directions. Red and blue lines represent opposite spin-polarized energy bands. Reproduced with permission of theoretical Ref.[34], copyright 2019, The American Physical Society.

*Non-magnetic materials with Rashba spin splitting:* Representative materials with large Rashba spin splitting are trigonal BiTeI[32], rhombohedral $\alpha$-SnTe[12], and quantum wires[42–44] and wells[45–48]. Fig. 3(D) shows the band structure of trigonal BiTeI with giant Rashba spin splitting at the band edges[32]. This spin splitting is enabled by the breaking of inversion symmetry $I$, as observed in the crystal structure shown in Fig. 3(C), and induced by the SOC. The magnitude of the strong Rashba effect is not well correlated with the atomic SOC strength[49]. Remarkably, the presence of energy band anti-crossing acts as a causal design principle of bulk materials with a large Rashba scale[49]. Based on this design principle, 34 strong Rashba compounds, such as trigonal PbS, trigonal $Sb_2Se_2Te$, and trigonal GeTe, associated with 165 weak Rashba compounds were predicted[49].

*Rashba spin splitting combined with topology or ferroelectricity:* Materials with Rashba spin splitting can have bifunctionality of other properties as the hallmark of large Rashba scale or the enabling symmetry condition is identical. For example: (i) Band anti-crossing as a hallmark of large Rashba scale also exists in all topological insulators, therefore those topological insulators with non-zero electric dipoles must have a strong Rashba effect[49]. An example of such topological Rashba materials is $Sb_2Se_2Te$, which has strong Rashba spin splitting and anti-crossing bands near the Fermi level. (ii) Broken inversion symmetry $I$ and non-zero dipoles in polar space groups as Rashba-enabling conditions are also required by ferroelectric functionality, therefore those Rashba materials that have a double potential well with two opposite-polarized degenerate ground states and a surmountable barrier are also ferroelectric materials[50]. Examples of such ferroelectric Rashba materials that were fabricated on the convex hull are $BrF_5$, GeTe, and $Sn_2P_2Se_6$. These materials are theoretically predicted to have (i) large Rashba spin splitting of 31, 142, and 67 meV, and (ii) small energy barrier of 98, 19, and 15 meV, respectively.

*Dresselhaus spin splitting combined with chirality:* Materials with Dresselhaus spin splitting can also have the bifunctionality of other properties. The ordinary Dresselhaus effect usually has auxiliary non-chiral symmetries. Huang et al. revealed another type of Dresselhaus effect in ferri-chiral systems, where both Dresselhaus spin splitting and chirality appear[11]. Moreover, this type of Dresselhaus spin splitting can be switched by the chirality[11], different from the ordinary Dresselhaus effect in zinc-blende materials like GaAs which are not switchable. An example of such ferri-chiral materials with the new type of Dresselhaus spin splitting is $NaCu_5S_3$, which has a low chiral phase transition barrier of 12 meV/f.u.[11].

## (A,2) Apparent SOC-induced spin splitting in non-centrosymmetric materials where antiferromagnetism is a passive background

The apparent SOC-induced spin splitting and spin polarization can also appear in AFM materials with high-*Z* elements where the antiferromagnetism is a passive background having no influence on the SOC-induced spin splitting and spin polarization[29,34,51–53]. The enabling symmetry is breaking $\Theta I$ while preserving $UT$. The enabling physical interaction is SOC, while the magnetic background breaking the time-reversal symmetry is not the essential physical interaction that causes spin splitting. This type broadens the apparent SOC-induced spin splitting and spin polarization from only NM crystals to some AFM systems.

*AFM Materials with SOC-induced spin splitting:* Representative materials with Rashba spin splitting nested in the AFM background are tetragonal $BiCoO_3$[34], orthorhombic $MnS_2$[29], and tetragonal $CaFe_2As_2$[29]. Fig. 3(E) shows the crystal structure of AFM tetragonal $BiCoO_3$[34]. As shown in Fig. 3(F), every band remains spin-degenerate when SOC is turned off and spin splitting only appears when SOC is turned on. This indicates that the spin splitting is uniquely induced by SOC. Notably, in the AFM $BiCoO_3$, the spin texture around Z point is also helical as the Rashba type[34]. This SOC-induced spin splitting of (A,2) subgroup is different from the SOC-independent non-relativistic spin splitting of (C,1) subgroup in both mechanism and features, which we will discuss in Section V.

# IV. Category (B) spin effects: Hidden SOC-induced spin polarization

*Hidden SOC-induced spin polarization inherited from the apparent counterpart:* Hidden SOC-induced spin polarization, shown as Category (B) in Fig. 1 and Fig. 2, inherits the local enabling symmetry and required physical interaction from the apparent counterpart with the difference that the physical properties of spin polarization are locally present but globally compensated. As shown in Figure 4(A, B), the enabling symmetry is breaking the inversion symmetry $I$ (combined with time-reversal symmetry $\Theta$ if applicable) in local environments while the inversion symmetry $I$ is preserved in the global system, and the induced physical interaction is SOC. Hidden SOC-induced spin polarization can exist in both NM and AFM materials.

*Local spin polarization projection from degenerate bands and experimental detectability:* Probing hidden spin polarization requires that the local polarization can be distinguished separately in real space. Otherwise, the global wavefunctions as a linear superposition of local wavefunctions will make the detection of local sectors undistinguished in real space and the local polarization will only be sensitive to the surface detection[54]. The requirement of local segregation of wavefunctions in real space is naturally satisfied in certain layered materials, such as 2H-$MoS_2$, where coupling between layers is often weak[55]. Or it can be satisfied by non-symmorphic symmetry, either two-fold screw axis or glide mirror plane[14,56], that enforces spin-sector locking on certain Brillouin Zone edges as exemplified by tetragonal $BaNiS_2$. The recently proposed hidden Zeeman-type spin polarization in NM materials (hexagonal $WSe_2$)[57], which has opposite collinear spin polarization in two local sectors at the Brillouin zone boundary, is a subtype of hidden Dresselhaus spin polarization. This spin texture at the Brillouin zone boundary is protected by non-symmorphic symmetries such as glide mirror plane or screw axis, which can even keep this Zeeman-type spin texture in small regions around time-reversal-invariant (TRI) points in non-centrosymmetric NM materials (hexagonal $BaBi_4O_7$)[57]. The experimental detection of hidden spin polarization can be achieved through (i) direct measuring of the segregated local spin polarization or (ii) indirect measuring other physical property that is apparent but dependent on the local spin polarization. The direct measurement (i) entails the introduction of a probe that can resolve the local sector or by introducing dynamic[58] or static perturbation that slightly breaks the global symmetry to induce the hidden effects. For example, one can detect hidden spin polarization via spin-angular-resolved-photoemission-spectroscopy (spin-ARPES) using a polarized probing beam that penetrates the material sample along the sector stacking direction[59–66]. The indirect measurement (ii) requires the identification of another spin-polarization-related property. For example, the hidden spin polarization in bilayer transition-metal chalcogenides can be measured by the optically pumped intrinsic circular polarized light, as the two layers emit light with the same handedness[63].

## (B,1) Hidden SOC-induced spin polarization in centrosymmetric non-magnetic materials

*Hidden Rashba and Dresselhaus spin polarization:* The inversion symmetry $I$ in centrosymmetric NM materials disallows the apparent Rashba and Dresselhaus spin splitting and spin polarization. However, if one projects quantum states into two non-centrosymmetric individual sectors, which are connected by the inversion symmetry $I$, non-zero spin polarization in subband states with opposite directions and same amplitude will arise in each sector. The local spin polarization in one local sector is fully compensated by the spin polarization from the other sector leading to zero net spin polarization in the global system. Hidden Rashba or Dresselhaus spin polarization, referred to as R2 and D2 corresponding to the apparent counterparts R1 or D1, is determined by non-centrosymmetric site point groups in the presence of global centrosymmetric point groups. If all atomic sites have non-centrosymmetric non-polar point groups ($D_2$, $D_3$, $D_4$, $D_6$, $S_4$, $D_{2d}$, $C_{3h}$, $D_{3h}$, $T$, $T_d$, $O$), the material will only produce the hidden Dresselhaus spin polarization. However, if at least one atomic site has polar point groups ($C_1$, $C_2$, $C_3$, $C_4$, $C_6$, $C_s$, $C_{2v}$, $C_{3v}$, $C_{4v}$, $C_{6v}$) with non-zero net dipoles, the material will also generate the hidden Rashba spin polarization.

*Materials with hidden Rashba spin polarization:* Representative materials of hidden Rashba spin polarization are tetragonal $BaNiS_2$[14–16] and tetragonal $LaOBiS_2$[12] as observed experimentally[35]. Fig. 4(D) shows that the local Rashba spin polarization dominates over the local Dresselhaus pattern in tetragonal $LaOBiS_2$[12]. As shown in Fig. 4(C), the crystal structure has the global space group of *P4/nmm* which possesses the inversion center, but the unit cell can be divided into three sectors, where two non-centrosymmetric $BiS_2$ layers are termed the $\alpha$-sector and $\beta$-sector and the central LaO 'blocking' layer has no impact on the energy spectrum near the Fermi level. The polar site point group $C_{4v}$ on Bi atoms results in a large dipole field of $1.8\times 10^5$ kV/cm, giving rise to the hidden Rashba spin polarization. The helical spin textures of $\alpha$-sector and $\beta$-sector have opposite directions and are fuly compensated by each other[12].

*Materials with hidden Dresselhaus spin polarization:* hidden Dresselhaus spin polarization proposed theoretically in Ref.[12], exhibits local spin texture but no spin splitting in the band structure. Representative materials of hidden Dresselhaus spin polarization are hexagonal NaCaBi[12], cubic Si[12], and cubic Ge[12]. Figure 4(F) shows the hidden Dresselhaus spin polarization in the hexagonal NaCaBi[12]. As shown in Fig. 4(E), the space group is $P6_3/mmc$ which possesses an inversion center, but the unit cell can be divided into two sectors of CaBi layers with a non-centrosymmetric non-polar site point group $D_{3h}$ on Ca and Bi atoms. Due to the absence of local dipoles, each of the two sectors independently generates local Dresselhaus spin polarization which is towards opposite directions and fully compensated by the global inversion symmetry[12].

**Figure 4. Hidden SOC-induced spin polarization [Category (B)].**

(A) Schematic of the centrosymmetric crystal structure with two non-centrosymmetric sectors.

(B) Schematic of the band structure with SOC-induced compensated spin polarization in local sectors.

(C) Crystal structure of NM centrosymmetric tetragonal $LaOBiS_2$. Red, green, purple, and orange atoms denote O, La, Bi, and S, respectively, with site point groups. Reproduced with permission of theoretical Ref.[12], copyright 2014, Springer Nature.

(D) Band structure and spin texture of NM centrosymmetric $LaOBiS_2$ with SOC-induced hidden spin polarization. Reproduced with permission of theoretical Ref.[12], copyright 2014, Springer Nature.

(E) Crystal structure of NM centrosymmetric hexagonal NaCaBi. Blue, red, and yellow atoms denote Ca, Bi, and Na, respectively, with site point groups. Reproduced with permission of theoretical Ref.[12], copyright 2014, Springer Nature.

(F) Band structure and spin texture of NM centrosymmetric NaCaBi with SOC-induced hidden spin polarization. Reproduced with permission of theoretical Ref.[12], copyright 2014, Springer Nature.

*Experimental observations related to hidden spin polarization:* Hidden spin polarizations in nonmagnetic centrosymmetric compounds are both theoretically and experimentally demonstrated to play an important role in producing several physical effects. Examples are: (i) Differential absorption of circularly polarized light was measured in $Li_2Co_3(SeO_3)_4$, originating from an interference between linear dichroism and linear birefringence[10]; (ii) Even integer quantum Hall effect in two-dimensional material due to hidden Rashba spin polarization[67], as demonstrated in $Bi_2O_2Se$[68]; (iii) Circular photocurrents without applying electric bias due to hidden spin polarization in transition-metal dichalcogenides such as hexagonal $MoTe_2$[69]; (iv) Transient spin polarization on a ultrafast (femtosecond timescales) in transition-metal dichalcogenides such as hexagonal $WSe_2$[58]; (v) Coexistence of superconductivity and local Rashba spin polarization in Bi-based cuprates[23,70].

### (B,2) Hidden SOC-induced spin polarization in antiferromagnetic materials where antiferromagnetism is a passive background

*Hidden Rashba and Dresselhaus spin polarization with background AFM:* Hidden Rashba and Dresselhaus spin polarization can also exist in AFM materials where the antiferromagnetism is a passive background (i.e., the antiferromagnetism is not the mechanism that produces spin polarization in local sectors). The enabling symmetry becomes more complex: (i) In the global system, both the combination of inversion symmetry $I$ and time-reversal symmetry $\Theta$ and the combination of spin-rotation symmetry $U$ and fractional-translation symmetry $T$ are preserved, i.e. globally SST-2 (see Table I); (ii) In the local sectors, the combination of inversion symmetry $I$ and time-reversal symmetry $\Theta$ is broken, but the combination of spin-rotation symmetry $U$ and fractional-translation symmetry $T$ is preserved, i.e. locally SST-3 (see Table I). The existence of this local spin polarization depends on the choice of sectors. So far, AFM materials showing this hidden effect have not been reported.

## V. Category (C) spin effects: Apparent SOC-independent spin splitting

*Historic development of non-relativistic spin splitting:* In 1964, S. I. Pekar and E. I. Rashba proposed a SOC-unrelated magnetic mechanism suggesting that there exists a spatially dependent periodic effective magnetic field *h(r)* which couples with the spin $\sigma$[71]; Such spin-lattice coupling may remove the spin degeneracy of energy bands. This idea remained dormant for over half a century due to the lack of guiding principles to search for target materials and the explicit form of this inhomogeneous magnetic field $h(r)$, until Yuan, Wang, Luo, E. I. Rashba, and Zunger first theoretically demonstrated this idea in the AFM compound $MnF_2$ providing specific enabling symmetry conditions[18]. This section will discuss the apparent SOC-independent spin splitting of Category (C) in Fig. 1 and Fig. 2. This type of spin splitting can exist in both AFM, FM, and ferrimagnetic materials, even when SOC is not considered.

## (C,1) Apparent non-relativistic spin splitting in antiferromagnetic materials

*Physics of non-relativistic spin splitting in AFM materials:* It was previously expected that electronic states would be spin-degenerate in the absence of SOC in AFM compounds. This is because it was intuitively assumed that the two sublattices with opposite magnetic moments will compensate each other, giving rise to spin degeneracy. Recently, the existence of spin-split antiferromagnets was proposed from computational studies: Yuan et al.[18] formulated in 2020 the enabling symmetry conditions for NRSS. As shown in Figure 5(A, B), apparent SOC-independent spin splitting, or “non-relativistic spin splitting”(NRSS) is enabled by breaking both the combination of inversion symmetry $I$ and time-reversal symmetry $\Theta$ (i.e. space-time-reversal symmetry) and the combination of spin-rotation symmetry $U$ and fractional-translation symmetry $T$ (i.e. translational-spin-rotation symmetry). This SOC-independent spin splitting can exist in low-*Z* compounds and even centrosymmetric systems, with giant value reaching several hundreds of *meV* compared to the Rashba and Dresselhaus SOC-induced spin splitting. Furthermore, Yuan et al. showed that using the configurations of these two enabling symmetry conditions and the magnetic order all the 1651 magnetic space group (MSG) can be classified into 7 spin-splitting prototypes (SST), summarized in Table I. This establishes the enabling symmetry conditions for different systems[29].

*AFM Materials with non-relativistic spin splitting:* Representative spin-split AFM materials are tetragonal $MnF_2$[18], orthorhombic $LaMnO_3$[29,72], cubic $NiS_2$[29], and rhombohedral $MnTiO_3$[29], tetragonal $KRu_4O_8$[30], tetragonal $V_2Te_2O$[73], tetragonal $RuO_2$[74], rhombohedral $Fe_2O_3$[75], hexagonal MnTe[20], and hexagonal CrSb[76]. Figure 5(D) shows the calculated large SOC-independent NRSS in tetragonal AFM $MnF_2$[18]. As shown in Fig. 5(C), the unit cell of AFM $MnF_2$ has two Mn atoms with opposite local magnetic moments. The calculated magnetic moments of 4.7 $\mu_B$ are on Mn atoms aligned along the [001] direction in good agreement with the neutron-scattering measurement of 4.6 $\mu_B$. Both the symmetries $\Theta I$ and $UT$ are violated. Apparent spin splitting arises in the path along $M$-$\Gamma$ and $A$-$Z$. The momentum-dependent apparent spin splitting is 204 meV between VB1 and VB2 along $A$-$Z$ and 297 meV between VB3 and VB4 along $M$-$\Gamma$ without SOC. When SOC is considered in the Hamiltonian, the resulting spin splitting can slightly change, as the magnitude of SOC-induced spin splitting is usually much smaller than that of NRSS.

# Category (C): Apparent SOC-independent spin splitting

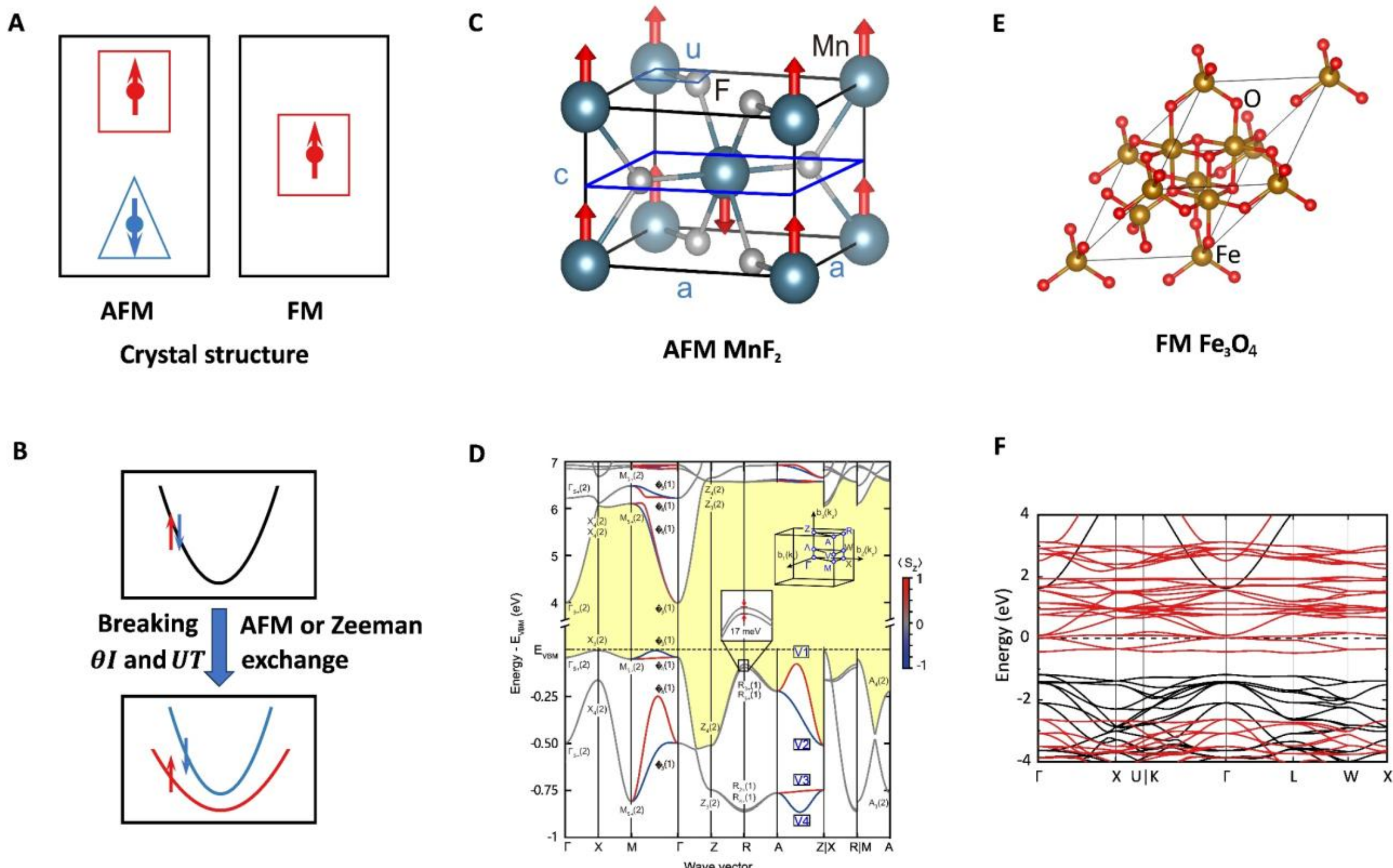

**Figure 5. Apparent SOC-independent spin splitting [Category (C)].**

(A) Schematic of the AFM or FM crystal structure with both $\Theta I$ and $UT$ symmetries violated. The red and blue arrows denote opposite local magnetic moments.

(B) Schematic of the band structure with SOC-independent spin splitting.

(C) Crystal structure of AFM tetragonal $MnF_2$. Blue and grey atoms denote Mn and F, respectively. Red arrows represent local magnetic moments. Reproduced with permission of theoretical Ref.[18], copyright 2020, The American Physical Society.

(D) Band structure of AFM tetragonal $MnF_2$ with SOC-independent spin splitting that is induced by AFM exchange interaction. Red and blue lines represent opposite spin-polarized energy bands. Reproduced with permission of theoretical Ref.[18], copyright 2020, The American Physical Society.

(E) Crystal structure of FM cubic $Fe_3O_4$. Brown and red atoms denote Fe and O, respectively.

(F) Band structure of FM cubic $Fe_3O_4$ with SOC-independent Zeeman spin splitting. Red and blue lines represent opposite spin-polarized energy bands.

*Subgroups of non-relativistic spin-split antiferromagnets and its connection to "altermagnets":* The spin-split AFM materials can be further classified into several subgroups, based on additional auxiliary spin-interconverting symmetries that potentially connect the two spin-opposite sublattices[21] and polarity. The auxiliary symmetries do not change the very existence of NRSS but give rise to various prototypical spin splitting and spin textures. The $\alpha$-type is the subgroup that has no spin-interconverting symmetry[77] although the chemical identities of the two opposite-spin ions are the same but has zero magnetization at zero temperature due to identical number of occupied spin-up and spin-down valence electrons which is "filling-enforced"[78,79], and almost zero magnetization under external perturbations. Spin splitting appears in the Brillouin-zone c*enter* (referred to as $\alpha$-type) due to the absence of auxiliary symmetry. An example of $\alpha$-type materials is $BiCrO_3$[21]. The $\alpha$-type AFM subgroup can be distinguished from the "compensated ferrimagnets" which also has zero magnetization at zero temperature and NRSS at Brillouin-zone center: The compensated ferrimagnets can have a large magnetization at either finite temperatures or when having different chemical magnetic species[21]. When the spin-interconverting symmetry is rotation, this subgroup is referred to as $\beta$-type. The $\beta$-type materials, such as $MnF_2$, have alternating spin polarization at rotation-connected *k*-paths, and were recently called "altermagnets"[30,31]. When the spin-interconverting symmetry is exclusively mirror reflection, this subgroup is referred to as $\gamma$-type. An example of $\gamma$-type materials is $MnTiO_3$[29]. Both $\alpha$- and $\beta$-type materials can be either polar or non-polar, whereas $\gamma$-type materials can only be polar. The polarity of these subgroups provides tunability for electric field switching of spin polarization.

## (C,2) Apparent Zeeman exchange interaction induced spin splitting in ferromagnetic materials

*Zeeman spin splitting in FM materials:* Zeeman effect was originally discovered in atoms by observing a spectral line separated into several components in the presence of a static magnetic field[80]. Expanding into solids, the FM compounds have a built-in intrinsic magnetic ordering, which provides Zeeman exchange interaction and gives rise to the Zeeman spin splitting, as a type of NRSS. The enabling symmetry for NRSS, i.e., violation of both (i) $\Theta I$ and (ii) $UT$, is satisfied in FM materials.

*FM Materials with non-relativistic Zeeman spin splitting:* Classic examples of FM elemental metals that have Zeeman spin splitting include Fe, Co, and Ni, all of which exhibit strong magnetic ordering. Such compounds are usually used to illuminate fundamental spintronic concepts as their spin-split electronic states can be directly observed and manipulated. Other representative FM compounds are spinel materials such as cubic $Fe_3O_4$[81], $HgCr_2Se_4$[82], and $CdCr_2Se_4$[83], and Heusler and half-Heusler materials such as cubic PtMnGa[84], $Co_2ZrAl$[85], and $Ni_2FeGa$[36]. Fig. 5(E) shows the crystal structure of cubic $Fe_3O_4$, where magnetic moments of 4.1 $\mu_B$ are on Fe atoms. Figure 5(F) shows the Zeeman spin splitting around all high-symmetric reciprocal directions. Notably, there exist opposite spin-polarized states near the Fermi energy. This is because the majority- and minority-spin states are shifted in energy under Zeeman exchange interaction, forming two distinct spin channels.

From an application perspective, Zeeman spin splitting in ferromagnets underpins the operation of technologies that form the backbone of modern magnetic memory and sensing, including magnetic tunnel junctions, giant magnetoresistance devices, and spin valves. Moreover, the interplay between Zeeman spin splitting and SOC can lead to additional phenomena such as magnetic anisotropy and anomalous Hall effect, further broadening the utility of FM compounds in both fundamental physics and technological innovation.

# VI. Category (D) spin effects: Hidden SOC-independent spin polarization

Hidden SOC-independent spin polarization is shown as Category (D) in Fig. 1 and Fig. 2. Like hidden SOC-induced spin polarization, the local spin polarization in symmetry-related sectors fully compensate each other leading to zero net spin polarization in the global system. Unlike the hidden SOC-induced spin polarization, physical mechanism that causes spin polarization is SOC-independent—coupling between magnetism and local crystal field, as shown in Figure 6(A, B). Consequently, the enabling symmetry condition also differs. In collinear antiferromagnets, this requires that (i) global system preserves $\Theta I$ or $UT$; (ii) local sectors can be AFM or FM sectors, violate $\Theta I$ and $UT$. When $S_I = \Theta I$ or $S_T = UT$ symmetry is not violated, the AFM material exhibits no NRSS and belongs to SST-1, SST-2, or SST-3 (see Table I). But the symmetries of the *local sectors* (spatially segregated atomic sites) in the unit cell satisfy the enabling symmetry condition ($S_I$, $S_T$) = (No, No), locally can be either SST-4 or SST-5 (see Table I), so that local SOC-independent spin polarization is allowed despite globally hidden. In this section, we will review recent efforts in searching for AFM materials having hidden non-relativistic spin polarization (NRSP) and demonstration of unique properties.

## (D,1) Hidden spin polarization without SOC in antiferromagnetic materials with antiferromagnetic local sectors

*Physics of hidden SOC-independent spin polarization in AFM materials with AFM local sectors:* In collinear AFM materials of SST-1, 2, 3 (see Table I), when the symmetries of the local sectors belong to SST-4, the material can locally possess SST-4-like properties, such as producing local NRSP that is globally hidden. Such local NRSP can be further divided into subgroups of $\alpha$-, $\beta$-, and $\gamma$ -subtypes, depending on the auxiliary spin-interconverting symmetries within the AFM sectors—analogous to the subgroups of SST-4 AFM as discussed in Sec. (C,1).

*AFM Materials with hidden SOC-independent spin polarization where local sectors are AFM:* Representative AFM materials are tetragonal $Ca_2MnO_4$[17,37], $La_2NiO_4$[17], and $MnS_2$[86], tetragonal $CoSe_2O_5$[87], tetragonal $Fe_2TeO_6$[88], tetragonal $K_2CoP_2O_7$[89], tetragonal $LiFePO_4$[90], tetragonal $Sr_2IrO_4$[91], tetragonal $SrCo_2V_2O_8$[92], Rhombohedral $Cr_2O_3$[93], and tetragonal $Cr_2SO$ bilayer[94]. As an example, Guo[94] extended the idea of hidden spin polarization from nonmagnetic materials to antiferromagnetic materials with $\Theta I$ symmetry, producing zero net spin polarization in total. Each inversion-partner sector possesses local non-relativistic spin polarization, dubbed "hidden altermagnetism"[94]. Recently this hidden altermagnetism was observed in the AFM $Cs_{1-\delta}V_2Te_2O$ in experiments by neutron diffraction and angle-resolved photoemission spectroscopy (ARPES)[95]. Figure 6(D) shows the SOC-independent hidden non-relativistic spin polarization in tetragonal AFM $Ca_2MnO_4$[17], where the magnetic moments align in the (001) direction. The global MSG is $I4_1'/a'cd'$ which preserves $\Theta I$ symmetry and violates $UT$ symmetry, but the unit cell can be divided into two magnetic sectors which keep the AFM order and violate both $\Theta I$ and $UT$ symmetries, as shown in Fig. 6(C). Each sector produces SOC-independent local spin polarization along the (001) direction that is fully compensated by the other sector[17].

*Bifunctionality of hidden SOC-independent spin polarization with AFM local sectors and multiferroicity:* Materials with hidden SOC-independent spin polarization with AFM local sectors, can have bifunctionality of other properties. For example, Matsuda et al.[86] showed that conventional antiferromagnets with a nonzero propagation vector (Q vector) bring about nontrivial symmetry breaking without lifting spin degeneracy but induce 'hidden altermagnetic spin splitting' with $\beta$-type or $\gamma$-type AFM local sectors in the electronic structure. The AFM compound $MnS_2$, which belongs to SST-3 (see Table I) where spin degeneracy cannot be lifted unless SOC is included, was found to have multiferroic properties and hidden SOC-independent spin polarization. Matsuda et al.[86] classifies collinear antiferromagnets with spin-interconverting symmetries into three types, including altermagnets ($\beta$-type or $\gamma$-type NRSS SST-4 AFM materials), $\Theta I$-symmetric magnets (SST-1 that have $\Theta I$ symmetry but no $UT$), and $Q$-magnets (SST-2 or SST-3 that have time-reversal symmetry and non-zero propagation vector $Q$), and concludes that $Q$-magnets can be polar possessing multiferroic properties that interact with the hidden SOC-independent spin polarization, such as nonlinear transport and optical activity.

## (D,2) Hidden spin polarization without SOC in antiferromagnetic materials with ferromagnetic local sectors

*Physics of hidden SOC-independent spin polarization in AFM materials with FM local sectors*: In collinear AFM materials of SST-1, 2, 3 (see Table I), when the symmetries of the local sectors belong to SST-5 (i.e., FM sectors), the material can locally possess SST-5-like properties, such as local Zeeman spin splitting and spin polarization.

## Category (D): Hidden SOC-independent spin polarization

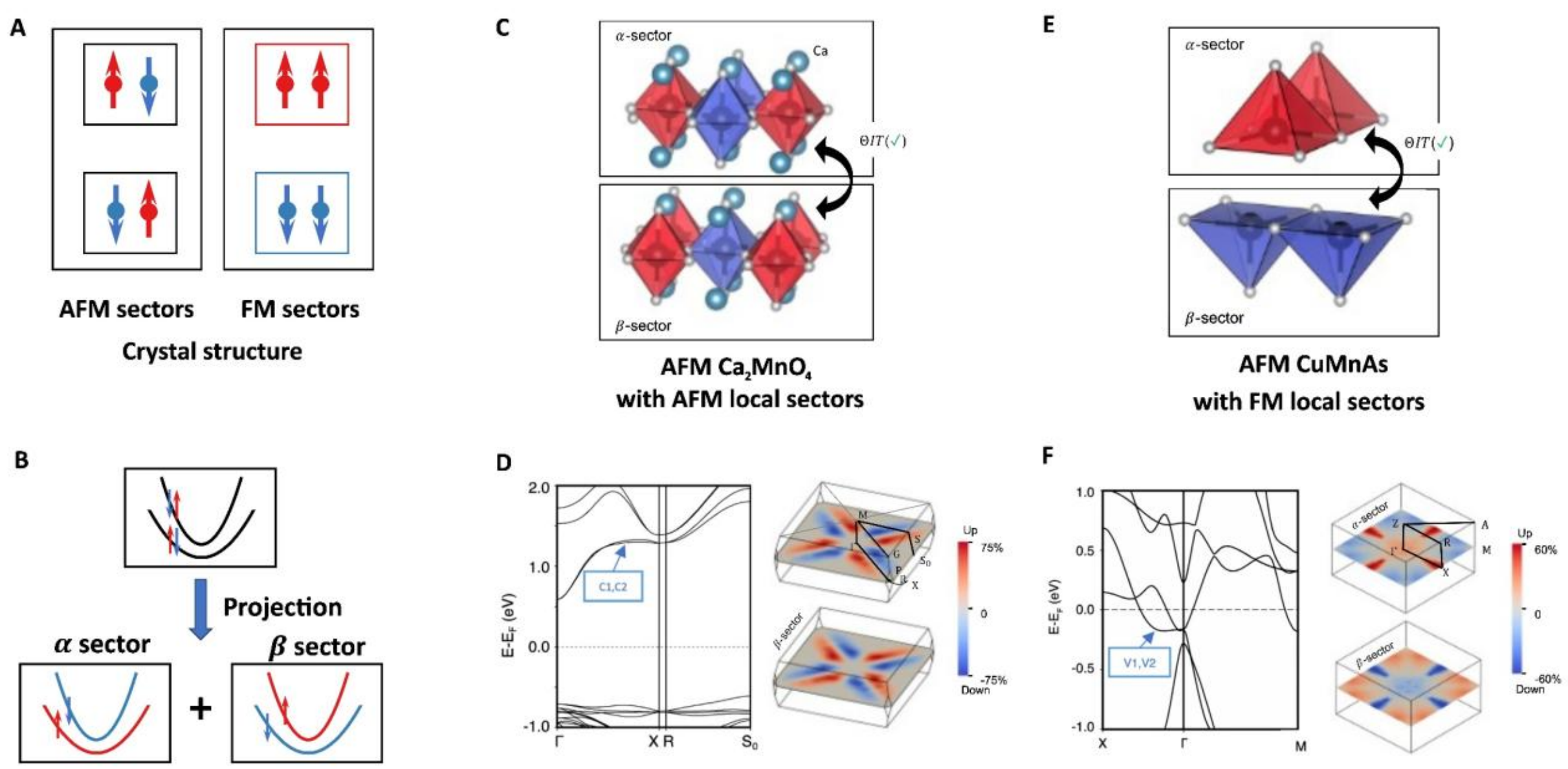

**Figure 6. Hidden SOC-independent spin polarization [Category (D)].**

(A) Schematic of the AFM crystal structure with two AFM or FM local sectors. The red and blue arrows denote opposite local magnetic moments.

(B) Schematic of the band structure with SOC-independent compensated spin polarization in magnetic local sectors.

(C) Crystal structure of AFM tetragonal $Ca_2MnO_4$ with two AFM local sectors connected by $\Theta I$ symmetry. Red and blue Mn-O octahedrons denote opposite local magnetic moments. Reproduced with permission (not required to obtain) of theoretical Ref.[17], copyright 2023, Springer Nature.

(D) Band structure and spin texture of AFM $Ca_2MnO_4$ with SOC-independent hidden spin polarization that is induced by AFM exchange interaction. Reproduced with permission (not required to obtain) of theoretical Ref.[17], copyright 2023, Springer Nature.

(E) Crystal structure of AFM tetragonal CuMnAs with two FM local sectors connected by $\theta IT$ symmetry. Red and blue Mn-As pentahedrons denote opposite local magnetic moments. Cu atoms are dismissed. Reproduced with permission (not required to obtain) of theoretical Ref.[17], copyright 2023, Springer Nature.

(F) Band structure and spin texture of AFM CuMnAs with SOC-independent hidden spin polarization that is induced by Zeeman exchange interaction. Reproduced with permission (not required to obtain) of theoretical Ref.[17], copyright 2023, Springer Nature.

*AFM materials with hidden SOC-independent spin polarization where local sectors are FM:* Representative AFM materials of this category are tetragonal CuMnAs[17,22,96], tetragonal $Mn_2Au$[97], hexagonal $FeBr_2$[17], hexagonal $CoCl_2$[17], hexagonal ErAuGe[17], trigonal MnSe bilayer[98]. As an example, Sheoran et al.[98] unveiled hidden Zeeman-type spin splitting in layered centrosymmetric antiferromagnets with asymmetric sublayer structures, i.e. bilayer MnSe, through first-principles simulations and symmetry analysis. They demonstrated that the degenerate states around specific *k* points spatially segregate on different sublayers forming $\theta I$-symmetric pairs. Figure. 6(F) shows the Zeeman exchange interaction induced hidden non-relativistic spin polarization in tetragonal AFM CuMnAs[17]. The magnetic moments align in the (010) direction. The global MSG is *Pm'mn* which preserves $\Theta I$ symmetry and violates $UT$ symmetry, but the unit cell can be divided into two magnetic sectors which have the FM order and violates both $\Theta I$ and $UT$ symmetries, as shown in Fig. 6(E). Each sector produces Zeeman-exchange-interaction induced local spin polarization along the (001) direction that is fully compensated by the other sector[17]. Note that CuMnAs is an AFM compound explored for current induced spin-orbit torque, but this property arises from its relativistic SOC effect, which is a hidden SOC-induced effect and differs from the hidden NRSP we show in Fig. 6(E).

Another case of hidden, SOC-independent spin polarization has been pointed out by Guo[99]. He considered two sectors, each having two compensated chemical magnetic species. Each of the two sectors satisfies individually the NRSS conditions. Because the magnetic moments of the individual magnetic species are compensated due to the conservation of number of occupied electrons, each local sector has zero magnetization despite having two different chemical magnetic species. This allows local spin splitting anywhere in the Brillouin zone, including the BZ center. By assumption, there are no auxiliary spin-interconverting symmetries in each sector (as in the $\alpha$ subgroup, see Ref. [21]). Also, by assumption, the two sectors are connected by either $\Theta I$ or $UT$ or both, therefore there can be no NRSS in the combined, two-sector system. This protects the compensation of spin polarization and spin degeneracy in the global AFM system. Guo referred this effect to "hidden fully compensated ferrimagnetism" and provided the trigonal bilayer AFM $CrMoC_2S_6$ as a representative material validated by DFT[99].

*Bifunctionality of hidden SOC-independent spin polarization with FM local sectors and nonreciprocal transport:* Materials with hidden SOC-independent spin polarization with FM local sectors can have the bifunctionality of other properties. For example, Chen et al.[22] predicts that macroscopic non-reciprocal transports induced by SOC-induced hidden spin polarization can coexist when coupled to another spatially distributed quantity. Hidden SOC-independent spin polarization with FM local sectors is a type of spatially distributed quantity in globally symmetry-unbroken antiferromagnets with symmetry-broken FM local sectors. An example of such AFM with nonreciprocal transport and hidden SOC-independent spin polarization and FM local sectors is CuMnAs as discussed in Ref. [22].

# VII. Electric tunability and switch of apparent and hidden spin splitting and polarization in antiferromagnets

In Sec. III-VI, we introduced various types of apparent and hidden spin splitting and polarization with enabling symmetries and physical interactions. The tunability as well as the ability to switch these spin effects will be important for both fundamental physics and potential device applications. Given that AFM compounds have the advantage of no stray field, rapid switching dynamics, and robustness against external magnetic perturbations, we will discuss in this section the electric tunability and switching of apparent and hidden spin splitting and polarization in AFM compounds. Switching can have different meanings in different contexts: (1) switching from apparent to apparent but reversed spin splitting; (2) switching from hidden to apparent; (3) switching of the Néel vector. We will introduce these three switching in the next three paragraphs highlighting electric tunability. Realizing the switching between apparent and hidden spin splitting and polarization in the same AFM compound will inspire new applications by controlling the magnetic configurations or crystal structures.

*Electric coupling between apparent NRSS and ferroelectric phases:* The apparent SOC-independent spin splitting [apparent NRSS, Category (C,1)] can be coupled to ferroelectricity[100]: The two ferroelectric phases (i) break the space-time-reversal and translation-spin-rotation symmetries thus enabling NRSS, (ii) have polar magnetic symmetry, (iii) are connected by space-time-reversal operation locating in two local potential wells (iv) with a low energy barrier between them, and (v) can be switched by an external electric field due to electric-magnetic coupling. Zhang et al. summarized all the polar magnetic point groups for collinear AFM compounds having NRSS[21], which can be a starting point in search of switchable ferroelectric SST-4 compounds. Notably, this electric switchable ferroelectric SST-4 phases can also be realized in two-dimensional systems, as discussed in Ref.[101].

*Electric coupling between hidden NRSP and ferroelectric ordering:* The hidden SOC-independent spin polarization [hidden NRSP, Category (D)] can also be coupled to ferroelectricity, while the apparent NRSS is coupled to antiferroelectricity[102]: Instead of the double potential wells of ferroelectric phases with one energy barrier, there are three phases with two energy barriers, in which two are ferroelectric non-SST-4 phases having hidden NRSP and the rest is antiferroelectric SST-4 phase having apparent NRSS and lowest total energy. The two non-SST-4 ferroelectric phases preserve either space-time-reversal symmetry or translation-spin-rotation symmetry or both leading to spin degeneracy throughout the whole Brillouin zone, while the SST-4 antiferroelectric phase breaks both spin-time-reversal and translation-spin-rotation symmetries producing NRSS. The non-SST-4 ferroelectric phases and the SST-4 antiferroelectric phase can be switched by an external electric field along different direction. This switching ability originates from the strong electric-magnetic coupling and low energy barrier that the non-SST-4 ferroelectric needs to overcome for switching.

*Electric switching by coupling to hidden SOC-induced spin polarization:* The two above cases of electrical switch utilizing apparent NRSS and hidden NRSP are not related to the relativistic SOC effect. However, the SOC effect, even in its hidden form (hidden SOC-induced spin polarization), plays an important role in AFM spintronic applications including non-reciprocal transport[22], current-induced spin current[96], anomalous Hall effect[103], and electrical tuned Néel vectors[104]. This is because the hidden SOC-induced spin polarization can be coupled to the external electric field or current, thus producing Néel-order spin-orbit torques (NSOTs). A famous example is the AFM compound tetragonal CuMnAs, which belongs to SST-1 (see Table I) that preserves the space-time-reversal symmetry and breaks the translation-spin-rotation symmetry. This compound has two FM $MnAs_5$ sectors that produce local NRSP and SOC-induced spin polarization, making it have both hidden NRSP and hidden SOC-induced spin polarization. The SOC effect is prominent as Mn states dominate the states near the Fermi level.

# VIII. A different type of hidden effects induced by farsightedness (hyperopia) of reduced local symmetries

Zunger[105] noted in 2002 that certain theories applied routinely in nanoscience, such as the truncated (small number of basic functions) $k \cdot p$ method are *farsighted* in the sense of not being able to resolve the correct atomistic symmetry of the studied object, replacing it by the average landscape of higher symmetry. This farsightedness, missing the correct, higher resolution local symmetry leads to failed prediction of a number of physical properties[105], such as the oscillating eigenvalues vs thickness of a thin film, the oscillating $\Gamma$-$X$ coupling in GaAs/AlAs quantum wells (QWs) and superlattices the heavy-hole-light-hole (HH-LH) coupling at Brillouin-zone center for common-atom and no-common-atom SLs, and *p*-level splitting in square-based pyramid quantum dots. In Sec. IV and Sec. VI of the current paper, we have discussed “hidden effects” where some of the enabling symmetry conditions in their apparent effects are nominally disallowed in the global systems but exist locally. In this section, inspired by the “farsightedness” scenario in nanostructures[105], we introduce another type of hidden spin effects that are induced by farsightedness of detailed reduced local symmetries in both real and reciprocal space. We point out a number of physical properties are masked by this symmetry farsightedness but are correctly revealed after replacing the farsightedness by higher resolution theories.

One example is the unusual spin texture patterns in NM spin-split crystals [Category (A)] that are not apparent from the space group as they are determined by the reciprocal space symmetry of the wavevector[39]. The crystallographic point group symmetry (CPGS) is equal to the wavevector point group symmetry (WPGS) at special wavevector such as the $\Gamma$ point but may also be different from the WPGS of certain wavevectors. The latter case will result in different spin texture pattern that is unexpected to appear in this crystal. For example[39], (i) rhombohedral GeTe with the polar CPGS $C_{3v}$ as a typical Rashba material has the helical spin texture pattern at the $Z$ point with the same WPGS $C_{3v}$, but exhibits a non-Rashba spin texture pattern at the $L$ point with the different WPGS $C_s$; (ii) Trigonal Te with chiral CPGS $D_3$ as a typical chiral material has the radial spin texture pattern at the $A$ point with the same WPGS $D_3$, but exhibits a non-radial spin texture pattern at the $M$ point with the different WPGS $C_2$; (iii) Cubic GaAs with the non-polar CPGS $T_d$ as a typical Dresselhaus material has the Dresselhaus spin texture pattern at the $\Gamma$ point with the same WPGS $T_d$, but exhibits a Rashba spin texture pattern at the $L$ point with the different polar WPGS $C_{3v}$. These (i) non-Rashba spin texture pattern in Rashba materials, (ii) non-chiral spin texture pattern in chiral materials, and (iii) Rashba spin texture pattern in Dresselhaus materials, reflect that the spin texture patterns are "hidden" in the CPGS enabled spin split materials. These hidden spin polarizations enabled by the local WPGS, as a new type of hidden effects, are not compensated by the global CPGS. Furthermore, they are pointed out to contribute significantly to the unconventional spin transport in NM crystals[106–108].

Another example is the heavy-hole-light-hole (HH-LH) mixing and the resulting linear Rashba spin splitting in the presence of an external electric field in zinc-blende and diamond structure quantum wells (QWs)[46,47,109]. It is well known that [001] GaAs/AlAs QWs have the $D_{2d}$ point group, under which HH and LH states have different irreducible representations (IREPs) at the $\Gamma$ point, prohibiting the HH-LH mixing. However, Ivchenko and Kaminski revealed that the HH-LH mixing is allowed by the reduced local interfacial symmetry $C_{2v}$[109]. Xiong et al. also demonstrated the existence of HH-LH mixing in Ge/Si QWs, which is the origin of linear Rashba spin splitting around the $\Gamma$ point in the presence of an external electric field[46,47]. The Ge/Si QWs with odd and even monolayers of wells and barriers have different global QW symmetries of $D_{2d}$ and $D_{2h}$, respectively. However, both have the same reduced local interfacial symmetry $C_{2v}$ enabling the HH-LH mixing and the resulting linear Rashba spin splitting. The local symmetry enabling certain important physical effects due to atomic arrangement at the interfaces of quantum structures, as "hidden" but not compensated by the global symmetry, is sometimes neglected by "farsightedness" and naively replaced the higher global symmetry, leading to ignorance of the corresponding physical effects[105].

# IX. Conclusions

In this Perspective, we show targeted physical properties of apparent and hidden effects with focus on spin splitting and spin polarization in different material platforms. We unify the apparent and hidden spin-related effects and classify them into four categories depending on their different enabling symmetries and underlying physical interactions, namely apparent SOC-induced spin splitting, apparent SOC-independent spin splitting, hidden SOC-induced spin polarization, and hidden SOC-independent spin polarization. We then discuss the electric tunability and switch of apparent and hidden spin splitting and spin polarization in AFM compounds. Furthermore, we extend the concept of hidden effects to "farsightedness" of local symmetries in both real space and reciprocal space that ignore the existence of physical properties. This Perspective paves the way to motivate the search of materials with targeted physical properties in different apparent and hidden physical effects.

## Declaration of Interests

The authors declare no competing interests.

## Acknowledgements

This work was supported by the National Science Foundation (NSF) DMR-CMMT Grant No. DMR-2113922. This work used resources of the Anvil system at Purdue University through Allocation No. PHY180030 from the Advanced Cyberinfrastructure Coordination Ecosystem: Services & Support (ACCESS) program, which is supported by National Science Foundation Grants No. 2138259, No. 2138286, No. 2138307, No. 2137603, and No. 2138296. We would like to thank Dr. Qihang Liu for valuable discussions.